\documentclass[11pt]{article}

\usepackage{axodraw4j}
\usepackage{amsfonts}
\usepackage{amsmath}
\usepackage{amssymb}
\usepackage{graphicx}
\usepackage{subfigure}
\usepackage{ccaption}

\captiondelim{. }

\def \be {\begin{equation}}
\def \ee {\end{equation}}
\def \bea {\begin{eqnarray}}
\def \eea {\end{eqnarray}}

\def \rr {\raise.35ex\hbox{\small $\prime$}\kern-.17em{\mbox{\large $\imath$}}}
\def \del {\partial}
\def \dels {\partial\kern-.5em / \kern.5em}
\def \pslash {p\kern-.5em / \kern.5em}
\def \kslash {k\kern-.5em / \kern.5em}
\def \As {{A\kern-.5em / \kern.5em}}
\def \Ds {D\kern-.7em / \kern.5em}
\def \Psib {{\overline \Psi}}
\def \a {\alpha}

\def \dag {\dagger}

\def \m {\mu}
\def \n {\nu}

\def \lam {\lambda}
\def \Lam {\Lambda}

\def \th {\theta}

\begin{document}
\begin{titlepage}

\begin{center}
\hfill hep-th/yymmnnn\\
\vskip .5in

\textbf{\Large
A UV completion of scalar electrodynamics}
\vskip .5in
{\large Pei-Ming Ho,$^a$ Xue-Yan Lin$^{b,}$
}

\vskip 15pt

{\small $^a$Department of Physics,
Center for Theoretical Sciences \\
and Leung Center for Cosmology and Particle Astrophysics, \\
National Taiwan University, Taipei 10617, Taiwan,
R.O.C.}\\
\vspace{3mm}
{\small $^b$Faculty of Law, University of Amsterdam, \\
Oudemanhuispoort 4-6 1012 CN  Amsterdam, the Netherlands.
}

\vskip .2in
\sffamily{
pmho@phys.ntu.edu.tw \\
xueyan.lin@msa.hinet.net}

\vspace{60pt}
\end{center}

\begin{abstract}

In previous works, 
we constructed UV-finite and unitary scalar field theories
with an infinite spectrum of propagating modes 
for arbitrary polynomial interactions.
In this paper, we introduce infinitely many massive vector fields
into a $U(1)$ gauge theory 
to construct a theory with UV-finiteness and unitarity.

\end{abstract}
\end{titlepage}
\setcounter{footnote}{0}

\section{Introduction}

UV divergences in quantum field theories can be 
regularized by introducing higher derivatives 
in the kinetic term so that the propagator 
approaches faster to $0$ than $1/k^2$ at large momenta $k$. 
Although this is usually done at the cost of unitarity, 
in Ref. \cite{PeiMingYiYa}, a UV-finite $\phi^4$ theory 
in four dimensions with a higher derivative correction 
to the propagator of the form 
\begin{equation}
f(k^2)=\sum_{n=0}^{\infty} 
\frac{c_n}{k^2+m_n^2} 
\quad\quad (c_n>0\quad \forall n)
\label{propagatorPP1}
\end{equation}
is proved to preserve unitarity. 
Due to the condition $c_n > 0$, 
Cutkosky's rules \cite{Cutkosky} ensure  
purturbative unitarity for generic Feynman diagrams. 

While there are an infinite number of poles at $k^2 = -m_n^2$
in the propagator,
this theory is also equivalent to a theory with 
an infinite number of scalar fields with masses $m_n$. 
If $m^2_n \gg m^2_0$ for all $n > 0$, 
the low energy behavior of this theory 
is approximated by an ordinary scalar field theory 
with a single scalar field with mass $m_0$. 
By fine-tuning the masses $m_n$ of the high-energy modes
and the coefficients $c_n$,
one can construct scalar field theories 
with arbitrary polynomial interactions 
in arbitrary even dimensions 
that are UV finite and unitary \cite{PeiMingXueYan}.

The purpose of this paper is to extend the tricks used 
to construct UV-finite and unitary scalar field theories 
to Abelian gauge theories.
We start by giving a comprehensive overview 
of our UV-finite and unitary scalar field theories
in this section.
Several new insights beyond earlier works
\cite{PeiMingYiYa,PeiMingXueYan} are included.
In the next section,
we show how the addition of an infinite tower of 
fine-tuned spectrum of massive vector fields 
can result in UV-finiteness for a $U(1)$ gauge theory 
while keeping the unitarity intact.
We make a few brief comments in the last section.

According to \cite{PeiMingYiYa}, 
in order to avoid UV divergence 
in the four dimensional $\phi^4$ theory, 
the following conditions are sufficient:
\begin{subequations}
\label{condition}
\begin{align}
&\sum_{n=0}^{\infty} c_n m_n^2=0, \\
&\sum_{n=0}^{\infty} c_n=0.
\end{align}
\end{subequations}
To satisfy these conditions in which 
an infinite sum of positive numbers vanishes, 
analytic continuation is applied. 
A famous example of analytic continuation 
often used in string theory textbooks
is the evaluation of the infinite sum
\begin{equation}
\sum_{n=1}^\infty n.
\end{equation}
One first defines the Riemann-Zeta function 
by a converging series 
\begin{equation}
\zeta(s) = \sum_{n=1}^{\infty} \frac{1}{n^s} \qquad (\Re(s) > 1),
\end{equation}
which analytically continues to $s=-1$ so that
\begin{equation}
\sum_{n=1}^\infty n\rightarrow \zeta(-1)=-\frac{1}{12}
\end{equation}
and the sum is identified with a negative number.
The physical reason for analytic continuation 
can be understood as the following \cite{PeiMingXueYan}. 
Due to the use of certain computational techniques 
or one's choice of formulation, 
the validity of some mathematical expressions may be restricted, 
but often the corresponding physical quantities 
can be well-defined with a larger range of validity.
Relying on the analyticity of the physical problem, 
analytic continuation allows us to retrieve 
the full range of validity of our results, 
even though the validity of derivations is more restricted.

A closely related theory of fermions was considered 
earlier by Itzhaki \cite{Itzhaki}. 
In his theory,
there are an infinite number of fermionic fields 
with a four-point interaction in four dimensions. 
The constraint on fermion masses is of the form
\begin{equation}
\sum_n c_n m^r_n=0,\qquad r=0,1,2,3.
\label{fermicond}
\end{equation}

In Ref. \cite{PeiMingXueYan},
we generalized his idea\footnote{
Our treatment of the analytic continuation is slightly 
different from that of Itzhaki in Ref. \cite{Itzhaki}.
} 
to $\phi^n$ theories in arbitrary even space-time dimensions. 
We adopted the same type of propagators 
and found that for any given $n$ 
and even space-time dimension $d$,
we only need the conditions 
\be
\sum_n c_n m_n^{2r}=0 \qquad \mbox{for} 
\quad r = 0, 1, \cdots, \frac{d-2}{2}.
\label{generalcond}
\ee
Remarkably this condition is independent of $n$. 
That is, the same propagator suits all interactions.
As in \cite{PeiMingYiYa}, 
a set of fine tuned parameters is required 
for analytic continuation to be applied. 
The following example was given in \cite{PeiMingXueYan}.
For arbitrary $d$, let 
\begin{subequations}
\begin{align}
c_n&=\biggl[
1+x_1(n+1)+x_2(n+2)(n+1)+\cdots \nonumber\\ 
&\quad+x_{d/2}(n+\frac{d}{2})(n+\frac{d}{2}-1)\cdots (n+1)
\biggr]e^{zn} \qquad (n\ge0,\qquad x_i\ge 0),\\
m^2_n&=e^{an}.
\end{align}
\label{example1}
\end{subequations}
We will set $z$ and $a$ to be positive, 
and we can always choose the coefficients $x_i$'s
such that all the conditions in (\ref{generalcond}) are satisfied.
Here is how this works.
Denote $\rho\equiv e^{z+ar}$ for convenience. 
First assuming $\rho<1$ and then applying analytic continuation 
to reach the region $\rho>1$, 
we obtain the sum $\sum c_nm^{2r}_n$ in the form
\begin{subequations}
\begin{align}
\sum_{n=0}^\infty c_nm_n^{2r}&
=\frac{1}{1-\rho}+
x_1\frac{d}{d\rho}\left(\frac{1}{1-\rho}\right)+
x_2\frac{d^2}{d\rho^2}\left(\frac{1}{1-\rho}\right)+
\cdots x_{d/2}\frac{d^{\frac{d}{2}}}{d\rho^\frac{d}{2}}
\left(\frac{1}{1-\rho}\right)\\
&=\frac{1}{\xi}+\frac{x_1}{\xi^2}+\frac{x_2}{\xi^3}+
\cdots\frac{x_{d/2}}{\xi^{d/2+1}}
\equiv h(\xi),
\end{align}
\end{subequations} 
where $\xi\equiv\frac{1}{1-\rho}$, 
which is negative definite when $\rho>1$.
By this method, we have sufficient parameters 
$\{x_1,x_2\cdots x_{d/2}\}$ 
to fix the roots of $\xi$ at desired positions, 
\be
- |\xi_r| = (1-\rho)^{-1} = (1-\exp(z+ar))^{-1}
\ee
for $r=0,1\cdots \frac{d-2}{2}$. 
These roots are all negative for $z, a > 0$.
We can find corresponding $x_i$'s 
by simply comparing the coefficients with the following equation
\begin{equation}
\xi^{d/2+1} h(\xi)=c(\xi+|\xi_1|)(\xi+|\xi_2|)\cdots(\xi+|\xi_{d/2}|),
\label{findx}
\end{equation}
where $c$ is an arbitrary real positive number.
Apparently all $x_i$'s are positive because no negative coefficients
appear in (\ref{findx}). 
This satisfies the condition in (\ref{propagatorPP1}) 
and prevents the violation of unitarity.

In the previous work \cite{PeiMingXueYan}
we checked that
the higher derivative theory 
with the propagator (\ref{propagatorPP1})
can be equivalently described as a theory with 
ordinary kinetic terms but infinitely many fields of masses $m_n$
with couplings dictated by the coefficients $c_n$.

There is an alternative interpretation of these theories
that was not discussed in earlier works.
If we expand the propagator in (\ref{propagatorPP1})
\begin{equation}
f(k^2)=\sum_n\frac{c_n}{k^2+m_n^2}
=\frac{\sum_n c_n}{k^2}-\frac{\sum_n c_nm_n^2}{k^4}
+\frac{\sum_n c_n m_n^4}{k^6}-\cdots,
\label{propagatorexpansion}
\end{equation}
with the condition (\ref{condition}) satisfied, 
the first two terms in the expansion vanish. 
Then we sum over the remaining terms and get
\begin{equation}
f(k^2)=\sum_n \frac{c_n m_n^2}{k^4}
\left(
\frac{m_n^2}{k^2}-\frac{m_n^4}{k^4}+\cdots
\right)
=\sum_n\frac{c_n m_n^4}{k^4(k^2+m_n^2)}.
\label{dualpropagator}
\end{equation}
Defining a theory by this propagator is another approach 
to define the perturbation theory. 
It appears as if every particle in the spectrum 
is allowed to move at the speed of light because 
the $k^2=0$ state satisfies their equation of motion. 
However, we will see below that 
the massless excitation is not physical.

\begin{figure}[t]
\centering
\includegraphics[scale=0.7]{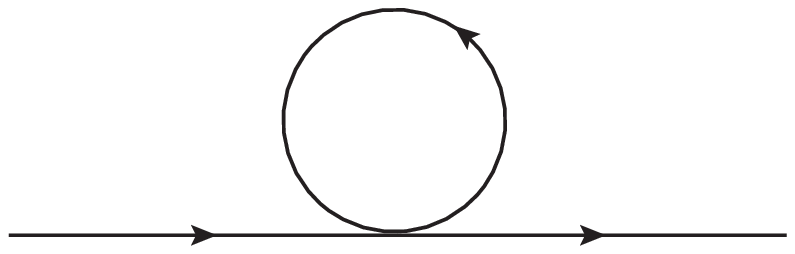}
\caption{}
\label{PhiPhi1loop}
\end{figure}

Using this propagator (\ref{dualpropagator}) to calculate 
the one-loop diagram for $\phi^4$ in 4 dimensions, 
we get
\begin{align}
\mathcal{M}&\propto
\sum_n\int d^4 l \frac{c_n m_n^4}{l^4(l^2+m_n^2)}
\nonumber\\
&\propto
\sum_n c_n m_n^2 
\left[
\log 1+
\log\left(\frac{0+m_n^2}{0}\right)
\right] \nonumber\\
&= \sum_n c_nm_n^2\log m_n^2 -\sum_n c_nm_n^2 \log 0,
\end{align}
with a potential infrared divergence in the second term. 
The IR divergence in fact does not exist 
because the coefficient $\sum_n c_n m_n^2$ vanishes 
as a result of the condition (\ref{condition}).
Therefore, the one-loop diagram is finite
\begin{equation}
\mathcal{M}\propto\sum_n c_nm_n^2\log m_n^2.
\end{equation}

The absence of IR divergence 
to be cancelled by soft massless particles implies that
there is no genuine massless excitation.
The conditions (\ref{condition}) remove
these massless propagating modes
by destructive interference among all particles $\phi_n$.
\footnote{
Here ``massless'' excitations does not refer to 
the propagating mode at $k^2=m^2_0$ 
if $m_0$ happens to be $0$.
}
This suggests that the massless mode does not propagate.
The absence of the massless propagating mode is 
crucial because the propagator $1/k^4$ violates unitarity.

Another way to see that $k^2=0$ is not a propagating mode
is to check that the propagator (\ref{dualpropagator}) 
does not diverge at $k^2 = 0$ because of (\ref{condition}). 
We rewrite the propagator (\ref{dualpropagator}) 
as a Laurent expansion of $k^2$, 
\begin{equation}
\sum_n\frac{c_n m_n^4}{k^4(k^2+m_n^2)}
=\frac{\sum_n c_nm_n^2}{k^4}-\frac{\sum_n c_n}{k^2}
+\sum_n\frac{c_n}{m_n^2}-\sum_n\frac{c_n k^2}{m_n^4}+\cdots,
\end{equation}
and find that it is actually a Taylor expansion
because the $1/k^4$ and $1/k^2$ terms both vanish.

Many features of the theory can be explained by 
the propagator (\ref{propagatorPP1}).
At high energy, the propagator behave as $1/k^6$. 
Therefore, the superficial divergence for a Feynman diagram 
with $L$ loops and $I$ internal lines in 4 dimensions is
of the order 
\begin{equation}
D=4L-6I.
\end{equation}
Because we always have $L\le I$ regardless of the details of interaction, the same conditions (\ref{condition}) for $\phi^4$ theory
is sufficient to guarantee UV finiteness for all $\phi^n$ theories
in 4 dimensions.

In general $d$ dimensions, 
$1/k^6$ at high energy does not guarantee $D < 0$, 
more conditions are required. 
In (\ref{propagatorexpansion}), 
the first $s+1$ terms vanish if we impose the conditions
\begin{equation}
\sum_n c_n=0,\qquad\sum_n c_n m_n^2=0,\cdots,
\qquad\sum_n c_n m_n^{2s}=0.
\label{condition_s}
\end{equation}
The propagator then behaves as $1/k^{4+2s}$ at high energy. 
To guarantee UV finiteness, 
we need the degree of superficial divergence to be negative, 
that is,
\begin{equation}
D=dL-(4+2s)I<0.
\end{equation}
Since $L\le I$, the condition $D<0$ is implied by
\begin{equation}
s>\frac{d-4}{2}.
\end{equation}
Therefore, 
in order for the conditions (\ref{condition_s}) to guarantee UV-finiteness,
the smallest possible value of $s$ is
\begin{equation}
s_{min}=\frac{d-2}{2}.
\end{equation}
That is, the conditions (\ref{generalcond})
ensure UV-finiteness in $d$ dimensions.

\section{UV-finite and unitary scalar QED}
\subsection{Lagrangian}

Our ansatz for the Lagrangian of 
a UV-finite and unitary $U(1)$ gauge theory is 
\begin{align}
\mathcal{L}&=\sum_n\frac{-1}{c_n}\left\{
\left[\left(\partial^\m-ie_n\sum_{a=0}^\infty A_a^\m\right)\phi_n\right]^\dagger
\left[\left(\partial_\m-ie_n\sum_{a=0}^\infty A_{a,\m}\right)\phi_n\right]
+m_n^2\phi_n^\dagger\phi_n\right\} \nonumber\\
&\;-\sum_a\frac{1}{b_a}\left(\frac{1}{4}F_a^{\m\n}F_{a,\m\n}
+\frac{1}{2}M_a^2 A_a^\m A_{a,\m}\right)
-V(\sum_n\phi_n^\dagger\phi_n). 
\label{Lagrangian}
\end{align}
We will assume the absence of scalar self interactions, 
that is, $V = 0$.
In Sec. \ref{Discussion} we will comment that if $V \neq 0$, 
UV divergence can not be avoided.
As in ordinary scalar QED, 
this theory contains two types of fields: 
the scalar fields $\phi_n$ and the vector fields $A_{a,\mu}$. 
One of the spin-one particles $A_{0,\mu}$ is massless $(M_0^2 = 0)$
and can be recognized as the photon. 
The gauge transformation rules of the fields are given by
\begin{subequations}
\begin{align}
\phi_n(x)&\rightarrow e^{-ie_n\Lambda(x)}\phi_n(x), \\
A_{0,\m}(x)&\rightarrow A_{0,\m}(x)-\partial_\m \Lambda(x),\\
A_{a,\m}(x)&\rightarrow A_{a,\m}(x),\quad n\ge 1.
\label{Transformation1}
\end{align}
\end{subequations}

Both considerations on unitarity and UV-divergence 
are centered around loop diagrams.
In the following we will focus on loop diagrams 
composed of the interaction vertices 
\be
e_n \phi_n^{\dag} \del_{\mu} \phi_n (\sum_a A_a^{\mu}), 
\qquad
e_n \del_{\mu} \phi_n^{\dag} \phi_n (\sum_a A_a^{\mu}), 
\qquad
e_n^2 \phi_n^{\dag} \phi_n (\sum_a A_a^{\mu})(\sum_b A_{b, \mu}).
\label{phiAinteraction}
\ee
See Fig. \ref{APhiPhi_Tree} and Fig. \ref{AAPhiPhi_Tree}.
These vertices do not mix $\phi_m$ with $\phi_n$ 
for $m\neq n$.
The transformations
\be
\phi_n \rightarrow e^{i\th_n} \phi_n
\label{phiphase}
\ee
are global symmetries of the theory,
unless the phases $\th_n$ are correlated by $\th_n \propto e_n$.
The particle number of each scalar $\phi_n$ is conserved.
Contrary to the scalar legs, 
the four point vertex $\phi\phi AA$
does not restrict the two vector particles to have the same index,
as shown in Fig. \ref{AAPhiPhi_Tree}. 
As a result, all vector internal lines 
are superpositions of all vector fields $A_{a,\mu}$.

\begin{figure}[!ht]
\centering
\subfigure[] {
\label{APhiPhi_Tree}
\scalebox{0.5}{
  \begin{picture}(247,233) (61,-37)
    \SetWidth{1.0}
    \SetColor{Black}
    \Line[arrow,arrowpos=0.5,arrowlength=5,arrowwidth=2,arrowinset=0.2](176,89)(272,169)
    \Line[arrow,arrowpos=0.5,arrowlength=5,arrowwidth=2,arrowinset=0.2](80,169)(176,89)
    \Photon(176,89)(176,-23){7.5}{6}
    \Text(110,110)[lb]{\LARGE{\Black{$p_1$}}}
    \Text(224,110)[lb]{\LARGE{\Black{$p_2$}}}
    \Text(200,80)[lb]{\LARGE{\Black{$ie_n(p_1+p_2)/b_a$}}}
    \Vertex(175,90){5.831}
    \Text(58,174)[lb]{\LARGE{\Black{$\phi_n$}}}
    \Text(273,175)[lb]{\LARGE{\Black{$\phi_n$}}}
    \Text(162,-42)[lb]{\LARGE{\Black{$A^\mu_a$}}}
  \end{picture}
}
}
\hspace{1cm}
\subfigure[] {
\label{AAPhiPhi_Tree}
\scalebox{0.5}{
  \begin{picture}(249,220) (76,-24)
    \SetWidth{1.0}
    \SetColor{Black}
    \Line[arrow,arrowpos=0.5,arrowlength=5,arrowwidth=2,arrowinset=0.2](95,172)(192,74)
    \Line[arrow,arrowpos=0.5,arrowlength=5,arrowwidth=2,arrowinset=0.2](193,76)(288,171)
    \Photon(96,-4)(190,73){7.5}{6}
    \Photon(271,-2)(192,76){7.5}{6}
    \Text(122,99)[lb]{\LARGE{\Black{$p_1$}}}
    \Text(236,99)[lb]{\LARGE{\Black{$p_2$}}}
    \Text(148,124)[lb]{\LARGE{\Black{$-2i e_n^2 g^{\mu\nu}/b_a$}}}
    \Vertex(190,79){5}
    \Text(73,175)[lb]{\LARGE{\Black{$\phi_n$}}}
    \Text(290,173)[lb]{\LARGE{\Black{$\phi_n$}}}
    \Text(75,-25)[lb]{\LARGE{\Black{$A^\mu_a$}}}
    \Text(267,-29)[lb]{\LARGE{\Black{$A^\nu_b$}}}
  \end{picture}
}
}
\caption{}
\end{figure}

The propagator of the $a$-th vector field $A_{a,\mu}$ is
\begin{equation}
\frac{b_a\left(g_{\mu\nu}
+\frac{k_\mu k_\nu}{M_a^2}\right)}{k^2+M_a^2}.
\end{equation}
Due to the extra piece $k_{\mu}k_{\nu}/M_a^2$ 
in the propagator, 
the perturbation theory is non-renormalizable 
by naive power counting.
But we will avoid the UV divergence by 
fine-tuned cancellation among all particles in the spectrum.
Since all internal lines for vector fields are given by
superpositions of all vector fields,
the effect of introducing the massive vectors $A_{a, \mu}$
is equivalent to replacing the photon propagator by
\begin{equation}
\Delta_{\mu\nu}(k)
=\frac{b_0\left(g_{\mu\nu}-(1-\xi)\frac{k_\mu k_\nu}{k^2}\right)}{k^2}
+\sum_{a=1}^{\infty}
\frac{b_a\left(g_{\mu\nu}+\frac{k_\mu k_\nu}{M_a^2}\right)}
{k^2+M_a^2},
\label{Photopropa}
\end{equation}
where $\xi$ is a free parameter depending on the choice of gauge.

It should be noted that an equivalent form of Lorentz gauge 
appears as the divergence of the free field equation of $A_{a,\mu}$,
\begin{eqnarray}
0&=&\partial_\nu
\left[\frac{-1}{b_a}(\partial_\mu F_a^{\mu\nu}-M_a^2 A_a^\nu)\right] \nonumber\\
&=&
\frac{1}{b_a}M_a^2 \partial_\nu A_a^\nu.
\label{MassiveLorentzgauge}
\end{eqnarray}
The physical meaning to this condition is that 
a spin-one particle has only three physical polarizations. 

The purpose of the rest of the section is to find the conditions 
on the parameters 
$c_n$, $e_n$, $m_n^2$, $b_a$ and $M_a^2$
such that this theory is UV-finite and unitary.
We will first examine the conditions on unitarity, 
and then study in detail the UV divergence of several loop diagrams 
before concluding on generic diagrams.

\subsection{Unitarity}

In the previous subsection, 
we introduced massive vectors into our theory. 
Now we have to check that those massive vectors 
do not  endanger unitarity. 

\begin{figure}[t]
\centering
\subfigure[] {
\label{Phi4_LoopAPhiAPhi_tt}
\includegraphics[scale=0.37]{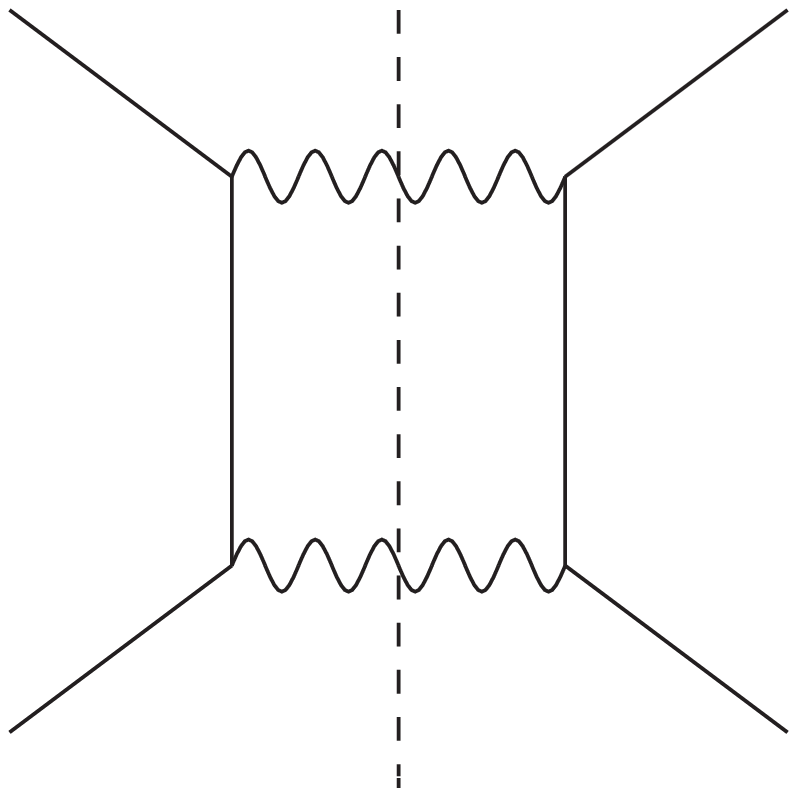}
}
\hspace{1cm}
\subfigure[] {
\label{Phi4_LoopAPhiAPhi_uu}
\includegraphics[scale=0.37]{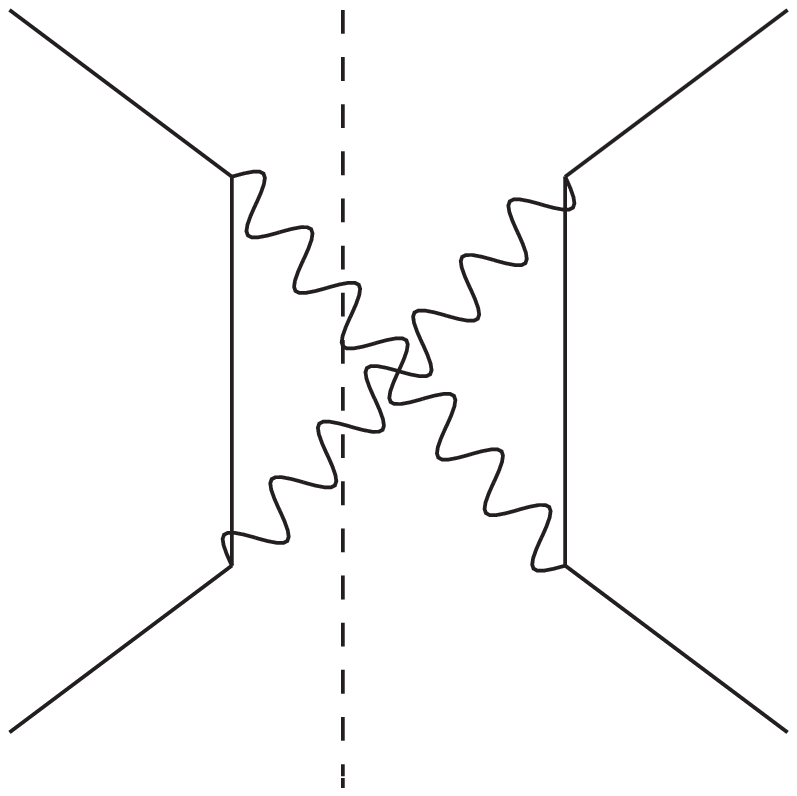}
}
\hspace{1cm}
\subfigure[] {
\label{Phi4_LoopAA}
\includegraphics[scale=0.37]{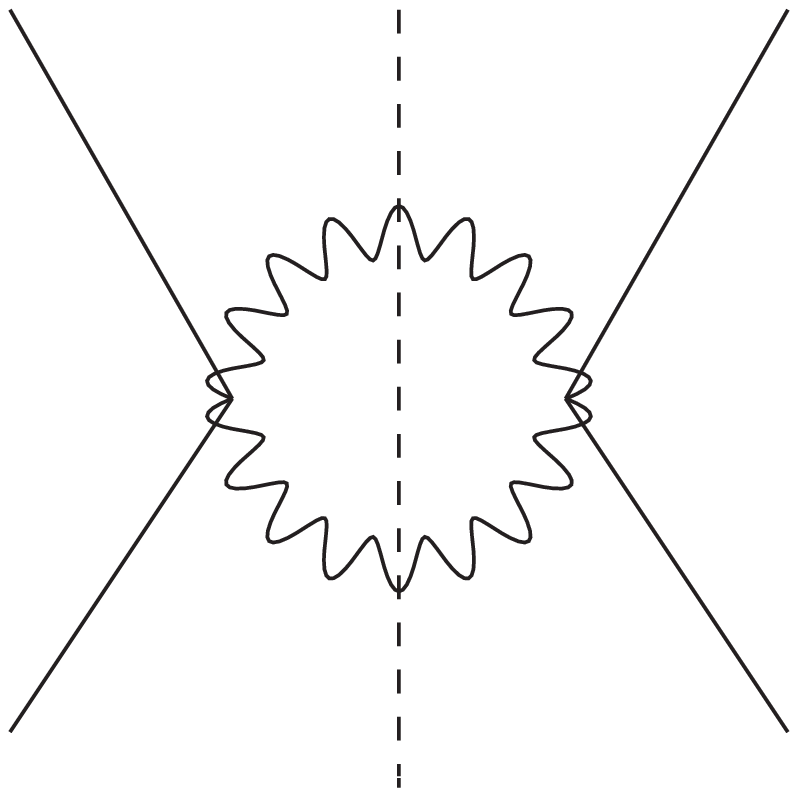}
}
\caption{}
\label{Phi4_unitarity}
\end{figure}
First we briefly review how unitarity survives 
in massless vector field theories. 
The requirement that the S-matrix is unitary implies that 
the scattering amplitude $T_{if}$ from the initial state $i$
to the final state $f$, defined as
\begin{equation}
S_{if}=\delta_{if}+i(2\pi)^4\delta(p_i-p_f)T_{if},
\end{equation}
satisfies the relation
\begin{equation}
T_{if}-T^*_{fi}=i\sum_n T_{in}T^*_{fn}(2\pi)^4\delta(p_i-p_f),
\label{Unitarity}
\end{equation}
where the sum is taken over all physical states 
consistent with all conservation laws. 
The left hand side can be calculated purturbatively 
by means of the so-called Landau-Cutkosky rules \cite{Cutkosky}: 
\begin{enumerate}
\item 
Cut through the diagram in all possible ways 
such that the cut propagators can simultaneously be put on shell.
\item 
Replace $1/(p^2+m^2+i\epsilon)$ by $-2\pi i \delta (p^2+m^2)$ 
for each cut propagator, then perform the loop integrals.
\item 
Sum over the contributions of all possible cuts.
\end{enumerate}

Using these rules, 
it is possible to check unitarity order by order 
in a perturbation theory. 
For example, 
Fig. \ref{Phi4_unitarity} shows three possible ways of the interaction 
$\phi\phi\rightarrow\phi\phi$ of order $O(e^4)$. 
The Landau-Cutkosky rules include the replacement of
the photon propagator in each internal line
\begin{equation}
\Delta_{\mu\nu}=\frac{g_{\mu\nu}-(1-\xi)\frac{k_\mu k_\nu}{k^2}}{k^2+i\epsilon} 
\rightarrow
-2\pi i \left[g_{\mu\nu}-(1-\xi)\frac{k_\mu k_\nu}{k^2}\right]\delta (k^2)\equiv -2\pi i G_{\mu\nu}\delta(k^2).
\label{polarsumvirtual}
\end{equation}
Therefore the left-hand side of (\ref{Unitarity}) is
\begin{equation}
\mathrm{Im}\mathit{T}=M^{\mu\rho}
G_{\mu\nu}G_{\rho\xi}
M^{\nu\xi},
\end{equation} 
where $M^{\mu\rho}$ is an on-shell amplitude of $\phi\phi\rightarrow AA$ which contains three diagrams (Fig. \ref{PPAA2}.).
\begin{figure}[t]
\centering
\subfigure[] {
\label{PhiPhiAA_Tree_t_2}
\scalebox{0.5}{
  \begin{picture}(188,231) (127,-27)
    \SetWidth{1.0}
    \SetColor{Black}
    \Line[arrow,arrowpos=0.5,arrowlength=5,arrowwidth=2,arrowinset=0.2](144,183)(208,135)
    \Line[arrow,arrowpos=0.5,arrowlength=5,arrowwidth=2,arrowinset=0.2](208,135)(208,39)
    \Line[arrow,arrowpos=0.5,arrowlength=5,arrowwidth=2,arrowinset=0.2](144,-9)(208,39)
    \Photon(208,135)(272,183){7.5}{4}
    \Photon(208,39)(272,-9){7.5}{4}
    \Text(124,183)[lb]{\Large{\Black{$\phi^+(p_1)$}}}
    \Text(126,-32)[lb]{\Large{\Black{$\phi^-(p_2)$}}}
    \Text(280,181)[lb]{\Large{\Black{$A(k_1)$}}}
    \Text(279,-31)[lb]{\Large{\Black{$A(k_2)$}}}
  \end{picture}
}
}
\hspace{1cm}
\subfigure[] {
\label{PhiPhiAA_Tree_u_2}
\scalebox{0.5}{
  \begin{picture}(188,231) (127,-27)
    \SetWidth{1.0}
    \SetColor{Black}
    \Line[arrow,arrowpos=0.5,arrowlength=5,arrowwidth=2,arrowinset=0.2](144,183)(208,135)
    \Line[arrow,arrowpos=0.5,arrowlength=5,arrowwidth=2,arrowinset=0.2](208,135)(208,39)
    \Line[arrow,arrowpos=0.5,arrowlength=5,arrowwidth=2,arrowinset=0.2](144,-9)(208,39)
    \Text(124,183)[lb]{\Large{\Black{$\phi^+(p_1)$}}}
    \Text(126,-32)[lb]{\Large{\Black{$\phi^-(p_2)$}}}
    \Text(280,181)[lb]{\Large{\Black{$A(k_1)$}}}
    \Text(279,-31)[lb]{\Large{\Black{$A(k_2)$}}}
    \Photon(208,135)(272,-9){7.5}{8}
    \Photon(208,39)(272,183){7.5}{8}
  \end{picture}
}
}
\hspace{1cm}
\subfigure[] {
\label{PhiPhiAA_Tree2_2}
\scalebox{0.5}{
  \begin{picture}(254,238) (127,-24)
    \SetWidth{0.5}
    \SetColor{Black}
    \Text(124,193)[lb]{\Large{\Black{$\phi^+(p_1)$}}}
    \Text(126,-25)[lb]{\Large{\Black{$\phi^-(p_2)$}}}
    \Text(346,190)[lb]{\Large{\Black{$A(k_1)$}}}
    \Text(346,-29)[lb]{\Large{\Black{$A(k_2)$}}}
    \SetWidth{1.0}
    \Line[arrow,arrowpos=0.5,arrowlength=5,arrowwidth=2,arrowinset=0.2](160,190)(256,94)
    \Line[arrow,arrowpos=0.5,arrowlength=5,arrowwidth=2,arrowinset=0.2](160,-2)(256,94)
    \Photon(256,94)(352,190){7.5}{7}
    \Photon(256,94)(352,-2){7.5}{7}
    \Vertex(255,95){4.123}
  \end{picture}
}
}
\caption{}
\label{PPAA2}
\end{figure}

On the right-hand side of (\ref{Unitarity}), 
the two diagrams obtained from cutting a diagram 
in Fig. (\ref{Phi4_unitarity}) are multiplied. 
A cut vector internal line becomes two vector external lines 
in the two diagrams of the $\phi\phi\rightarrow AA$ process.
The summation over polarizations of a pair 
of vector external lines
contributes the factor
\begin{equation}
\sum_{\lambda=1,2}\epsilon^*_\mu(\lambda,k)\epsilon_\nu(\lambda,k)
\equiv - E_{\mu\nu}.
\end{equation}
The sum only includes physical polarizations,
so
\begin{equation}
E_{\mu\nu}
= g_{\mu\nu} - \frac{k_\mu n_\nu+k_\nu n_\mu}{n\cdot k}
+ \frac{n^2 k_\mu k_\nu}{(k\cdot n)^2},
\label{polarsum}
\end{equation}
where $n_\nu$ is an arbitrary vector satisfying
\begin{equation}
k\cdot n\neq0,
\qquad 
n\cdot\epsilon(\lam, k)=0,
\end{equation}
Compare (\ref{polarsumvirtual}) and (\ref{polarsum}). 
Their difference is that the unphysical polarizations 
are included in $G_{\mu\nu}$ but not in $E_{\mu\nu}$. 
However, 
in ordinary QED there is no problem due to the fact that 
photons always couple to conserved currents
\begin{equation}
k_\mu M^{\mu\nu}=0.
\label{Ward}
\end{equation}
Therefore,
\begin{equation}
M^{\mu\rho}G_{\mu\nu}G_{\rho\xi}M^{\nu\xi}
=M^{\mu\rho}E_{\mu\nu}E_{\rho\xi}M^{\nu\xi}
=M^{\mu\rho}g_{\mu\nu}g_{\rho\xi}M^{\nu\xi}
\end{equation}
and unitarity is preserved.

In massive vector theories, 
the Ward-identity (\ref{Ward}) no longer applies. 
Yet the situation is actually simpler. 
The sum over physical polarizations is
\begin{equation}
\sum_{\lambda=1,2,3}\epsilon_\mu(\lambda,k)\epsilon^*_\nu(\lambda,k)
=g_{\mu\nu}+\frac{k_\mu k_\nu}{M^2},
\label{pol-sum}
\end{equation}
which is exactly the numerator of the massive vector boson propagator. 
It means that there is no unphysical polarizations in the internal lines. 
In massive theories, the number of physical polarizations is three, 
and it is exactly the number of degrees of freedom of spin-one particles. 
On the other hand, for a massless vector field, 
in order to obtain a well-defined propagator, 
one introduces a gauge-fixing term of the form 
\begin{equation}
\frac{-1}{2\xi}\Bigl(\partial_\mu A^\mu\Bigr)^2.
\end{equation}
In the propagator the number of polarizations is 4, 
but there are only 2 physical polarizations. 
Therefore, we need Ward-identities 
to ensure that unphysical intermediate states decouple.

A well known theory containing massive vector bosons 
is the intermediate vector boson (IVB) model of weak interaction.
In this theory heavy vector bosons $W^\pm$ are introduced 
to tame the high energy behavior of beta decay
in Fermi's four-fermion current-current model. 
But diagrams with external $W^\pm$ bosons
\cite{Aitchison,Gastmans},
\begin{figure}[t]
\centering
\includegraphics[scale=0.5]{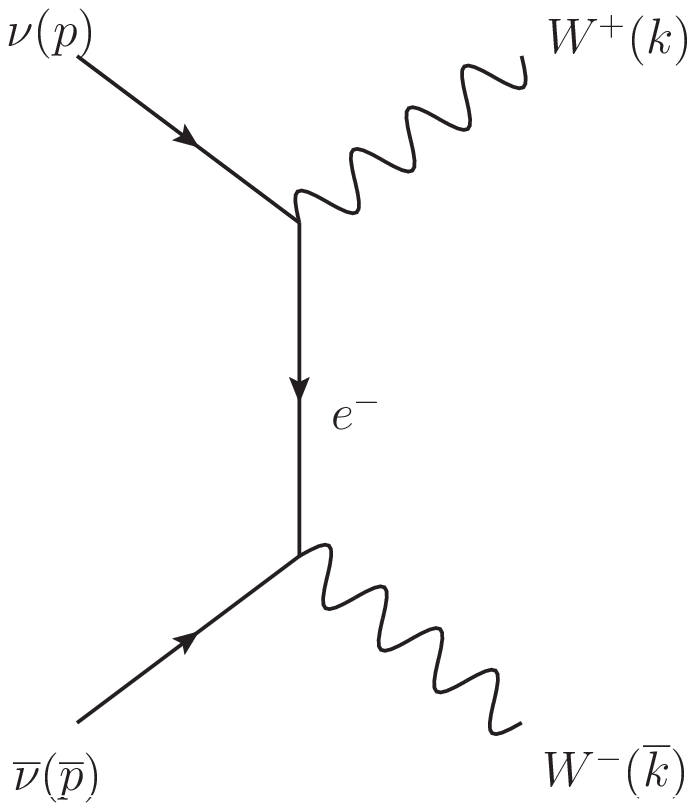}
\caption{}
\label{nunuWW_Tree}
\end{figure}
e.g. Fig. \ref{nunuWW_Tree},
still suffer high energy divergence.
At the leading order, the cross section is
\cite{Aitchison,Gastmans}
\begin{equation}
\frac{d\sigma}{d\Omega}=\frac{\sum_{pol}|M_t|^2}{64\pi^2s}
\simeq\frac{G^2E^2\sin^2\theta}{8\pi^2}.
\end{equation}
When the energy is large enough, 
the total probability of the scattering is larger than 1, 
and unitarity is violated, 
in a sense different from our discussions above.

The murderer of unitarity here 
is the longitudinal mode of $W^\pm$ bosons.
Recall the polarization sum (\ref{pol-sum}).
The second term in (\ref{pol-sum}), 
which results in bad high-energy behaviors, 
comes from the longitudinal mode. 

\begin{figure}[!ht]
\centering
\subfigure[] {
\label{nunuWW_Tree_Z}
\includegraphics[scale=0.5]{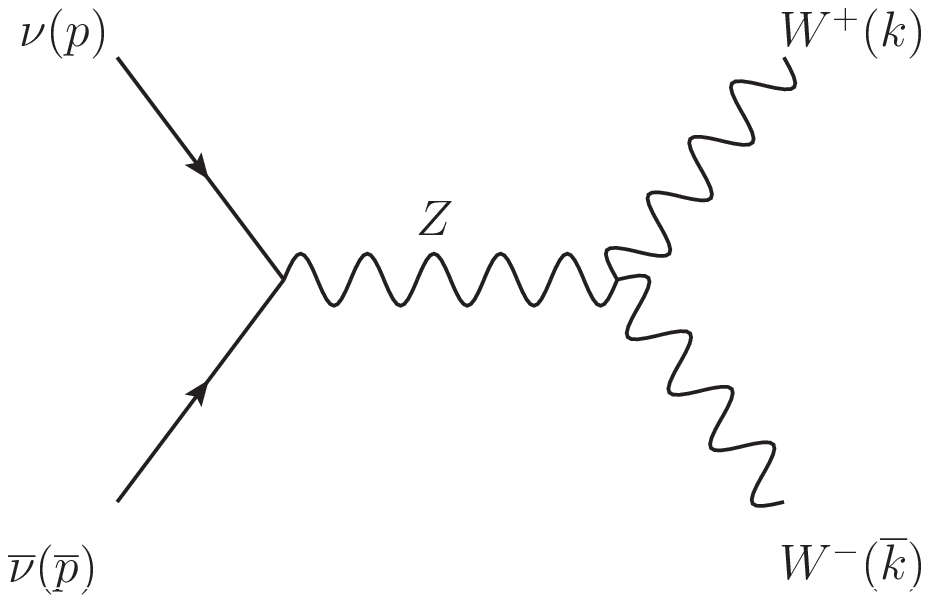}
}
\hspace{1cm}
\subfigure[] {
\label{nunuWW_Tree_u}
\includegraphics[scale=0.5]{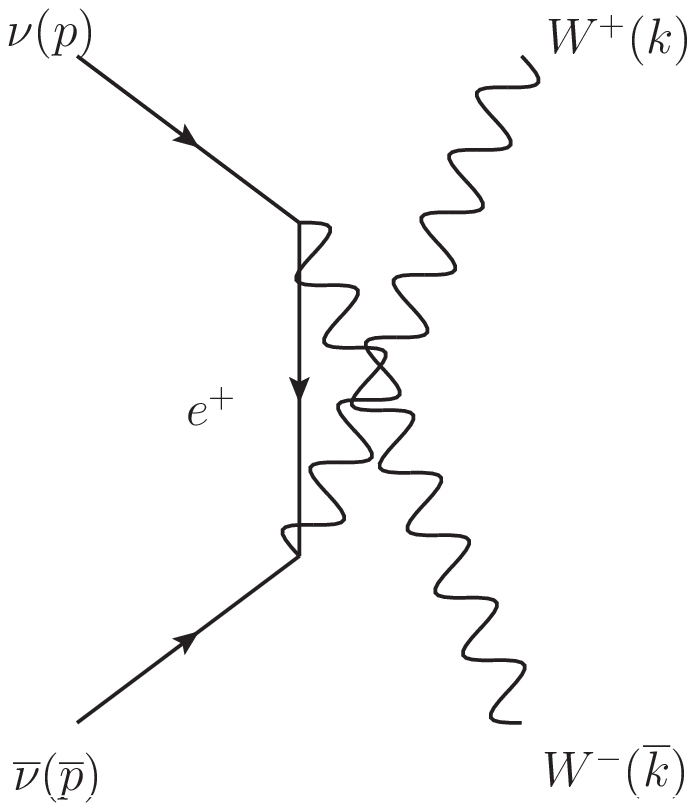}
}
\caption{}
\label{nunuWW_Tree_u+19}
\end{figure}

To cancel this $E^2$ term, 
destructive interference in s-channel (Fig \ref{nunuWW_Tree_Z}) 
and/or u-channel (Fig \ref{nunuWW_Tree_u}) is a possible way out. 
This motivated the introduction of the Z boson. 

In our theory, we will also rely on the destructive interference 
among diagrams with the same external legs. 
For example, the pair annihilation 
$\phi^+_n\phi^-_n\rightarrow A_a A_b$ 
contains three diagrams at the tree level 
(Fig. \ref{PhiPhiAA_Tree_t+21+22}),
\begin{figure}[t]
\centering
\subfigure[] {
\label{PhiPhiAA_Tree_t}
\scalebox{0.5}{
  \begin{picture}(188,231) (127,-27)
    \SetWidth{1.0}
    \SetColor{Black}
    \Line[arrow,arrowpos=0.5,arrowlength=5,arrowwidth=2,arrowinset=0.2](144,183)(208,135)
    \Line[arrow,arrowpos=0.5,arrowlength=5,arrowwidth=2,arrowinset=0.2](208,135)(208,39)
    \Line[arrow,arrowpos=0.5,arrowlength=5,arrowwidth=2,arrowinset=0.2](144,-9)(208,39)
    \Photon(208,135)(272,183){7.5}{4}
    \Photon(208,39)(272,-9){7.5}{4}
    \Text(124,183)[lb]{\Large{\Black{$\phi^+_n(p_1)$}}}
    \Text(126,-32)[lb]{\Large{\Black{$\phi^-_n(p_2)$}}}
    \Text(280,181)[lb]{\Large{\Black{$A_a(k_1)$}}}
    \Text(279,-31)[lb]{\Large{\Black{$A_b(k_2)$}}}
  \end{picture}
}
}
\hspace{1cm}
\subfigure[] {
\label{PhiPhiAA_Tree_u}
\scalebox{0.5}{
  \begin{picture}(188,231) (127,-27)
    \SetWidth{1.0}
    \SetColor{Black}
    \Line[arrow,arrowpos=0.5,arrowlength=5,arrowwidth=2,arrowinset=0.2](144,183)(208,135)
    \Line[arrow,arrowpos=0.5,arrowlength=5,arrowwidth=2,arrowinset=0.2](208,135)(208,39)
    \Line[arrow,arrowpos=0.5,arrowlength=5,arrowwidth=2,arrowinset=0.2](144,-9)(208,39)
    \Text(124,183)[lb]{\Large{\Black{$\phi^+_n(p_1)$}}}
    \Text(126,-32)[lb]{\Large{\Black{$\phi^-_n(p_2)$}}}
    \Text(280,181)[lb]{\Large{\Black{$A_a(k_1)$}}}
    \Text(279,-31)[lb]{\Large{\Black{$A_b(k_2)$}}}
    \Photon(208,135)(272,-9){7.5}{8}
    \Photon(208,39)(272,183){7.5}{8}
  \end{picture}
}
}
\hspace{1cm}
\subfigure[] {
\label{PhiPhiAA_Tree2}
\scalebox{0.5}{
  \begin{picture}(254,238) (127,-24)
    \SetWidth{0.5}
    \SetColor{Black}
    \Text(124,193)[lb]{\Large{\Black{$\phi^+_n(p_1)$}}}
    \Text(126,-25)[lb]{\Large{\Black{$\phi^-_n(p_2)$}}}
    \Text(346,190)[lb]{\Large{\Black{$A_a(k_1)$}}}
    \Text(346,-29)[lb]{\Large{\Black{$A_b(k_2)$}}}
    \SetWidth{1.0}
    \Line[arrow,arrowpos=0.5,arrowlength=5,arrowwidth=2,arrowinset=0.2](160,190)(256,94)
    \Line[arrow,arrowpos=0.5,arrowlength=5,arrowwidth=2,arrowinset=0.2](160,-2)(256,94)
    \Photon(256,94)(352,190){7.5}{7}
    \Photon(256,94)(352,-2){7.5}{7}
    \Vertex(255,95){4.123}
  \end{picture}
}
}
\caption{}
\label{PhiPhiAA_Tree_t+21+22}
\end{figure}
the amplitude is 
\begin{equation}
\mathcal{T}=-\frac{e_n^2}{c_n}
\left[
\frac{4(p_1\cdot\epsilon_1)(p_2\cdot\epsilon_2)}{m_n^2-t}
+\frac{4(p_1\cdot\epsilon_2)(p_2\cdot\epsilon_1)}{m_n^2-u}
+2(\epsilon_1\cdot\epsilon_2)
\right],
\label{PPAA}
\end{equation}
where
\begin{equation}
t=-(-p_1-k_1)^2,
\qquad
\mbox{and}
\qquad
u=-(-p_1-k_2)^2.
\end{equation}
Here the incoming momenta are defined 
to have negative energy and the relation of 
energy momentum conservation is 
\begin{equation}
p_1+p_2+k_1+k_2=0.
\end{equation}
Because we only worry about the $E^4$ term in $\mathcal{|T|}^2$, 
we focus our attention on leading order terms in the expansion
\begin{subequations}
\begin{align}
\frac{1}{m_n^2-t}&=
\frac{1}{m_n^2+(p_1+k_1)^2}
=\frac{1}{2(p_1\cdot k_1)-M_a^2}
=\frac{1}{2p_1\cdot k_1}[1+\mathcal{O}(1/E^2)], \\
\frac{1}{m_n^2-u}&=
\frac{1}{m_n^2+(p_1+k_2)^2}
=\frac{1}{2(p_1\cdot k_2)-M_b^2}
=\frac{1}{2p_1\cdot k_2}[1+\mathcal{O}(1/E^2)],
\end{align}
\end{subequations}
and the longitudinal polarizations are
\begin{subequations}
\begin{equation}
\epsilon^\mu_1\rightarrow \frac{k_1^\mu}{M_a},
\qquad
\epsilon^\mu_2\rightarrow \frac{k_2^\mu}{M_b}.
\end{equation}
\end{subequations}
Using these relations, 
the high energy behavior of the amplitude in (\ref{PPAA}) 
is largely simplified
\begin{equation}
\mathcal{T} \rightarrow-\frac{e_n^2}{c_n}
\left[
\frac{2(p_2\cdot k_2)}{M_a M_b}
+\frac{2(p_2\cdot k_1)}{M_a M_b}
+2\frac{(k_1\cdot k_2)}{M_a M_b}
\right]
= \frac{- e_n^2}{c_n}\frac{M_a^2+M_b^2}{M_a M_b}.
\label{PPAAhigh}
\end{equation}
The term of order $E^2$ is suppressed due to the identity
\be
s+t+u = 2m_n^2 + M_a^2 + M_b^2.
\ee
We see in (\ref{PPAAhigh}) the destructive interference 
among three diagrams. 
Unitarity is safe at high energies.

\begin{figure}[ht!]
\centering
\includegraphics[scale=0.5]{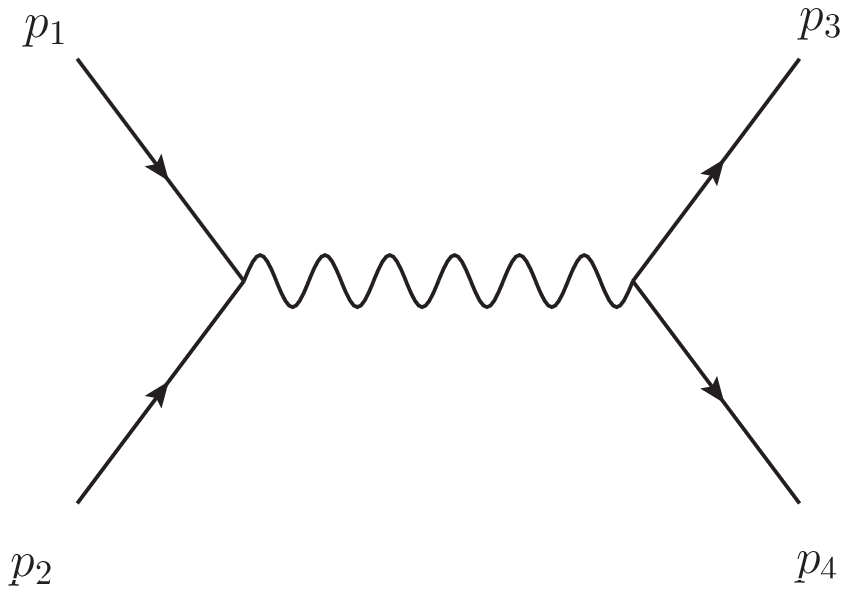}
\caption{}
\label{Phi4_Tree_A}
\end{figure}

Next we consider 
the $\phi^+_m\phi^-_m\rightarrow \phi^+_n\phi^-_n$ diagram 
(Fig. \ref{Phi4_Tree_A}).
The amplitude is 
\begin{equation}
\mathcal{M}_{mmnn}=
-ie_m(p_1-p_2)^\mu
\sum_a\frac{b_a(g_{\mu\nu}+\frac{(p_1+p_2)_\mu(p_3+p_4)_\nu}
{M_a^2})}{M_a^2-s}
(-ie_n)(p_3-p_4)^\nu.
\label{PPPP}
\end{equation}
Although the term $k^\mu k^\nu/M^2$ in the propagator 
leads to a term of 4th order in energy,
it vanishes because
\begin{equation}
(p_1-p_2)^\mu(p_1+p_2)_\mu(p_3+p_4)_\nu(p_3-p_4)^\nu
=(p_1^2-p_2^2)(p_3^2-p_4^2)=0.
\end{equation}
Instead of imposing the equation of motion on (\ref{PPPP}), 
there is another way to control the $k^\mu k^\nu$ term in propagator. 
Take the $1/E$ expansion of (\ref{PPPP}), 
\begin{equation}
\mathcal{M}_{mmnn}=
-ie_m(p_1-p_2)^\mu
\sum_a\left[-b_a\frac{(p_1+p_2)_\mu(p_3+p_4)_\nu}
{M_a^2 \, s}+\mathcal{O}(1/E^2)\right]
(-ie_n)(p_3-p_4)^\nu,
\label{PPPPhighen}
\end{equation}
the leading order term vanishes if the condition 
\begin{equation}
\sum_{a=1}\frac{b_a}{M_a^2}=0
\end{equation}
is imposed.
Although this is not necessary here, 
we will meet the same condition in the next subsection 
in order for certain loop diagrams to be finite.

In general, the potential violation of unitarity at high energies
due to massive vector bosons 
has its origin in the $k_{\mu}k_{\nu}/M^2_a$ term
in the propagator.
The same term is also responsible for potential nonrenormalizability.
Our basic idea is to fine-tune the parameters $b_a, M_a^2$
to achieve destructive interference among all massive vector bosons,
such that the vector propagator is sufficiently well-behaved 
at high energies for the theory to be UV-finite and unitary.

\subsection{UV divergences in loop diagrams}

In this subsection we calculate some one-loop diagrams, 
and show that UV divergences are canceled in every diagram
if certain conditions on the parameters of the theory are matched.
Then we discuss the cancellation in generic Feynman diagrams.

\subsubsection{Self-energy of $\phi$}

\begin{figure}[!ht]
\centering
\subfigure[] {
\label{PhiPhi_Loop_A}
\includegraphics[scale=0.5]{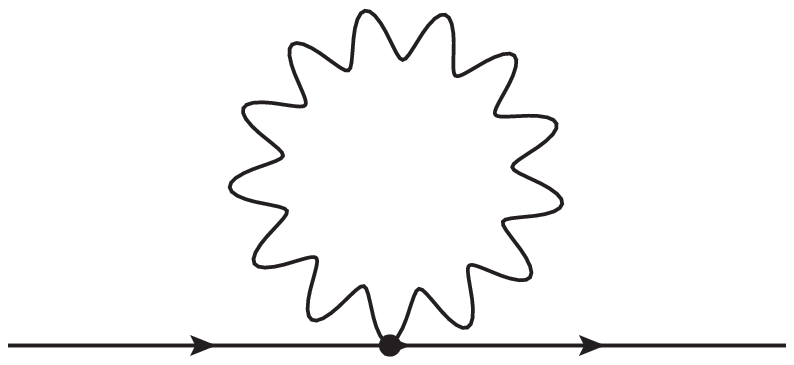}
}
\hspace{1cm}
\subfigure[] {
\label{PhiPhi_Loop_APhi}
\includegraphics[scale=0.5]{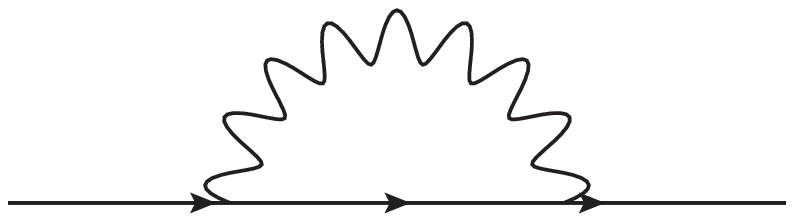}
}
\caption{}
\label{PhiPhi_Loop}
\end{figure}

We first calculate the self-energy diagram for the scalars. 
The two external lines must have the same index $n$. 
There are two diagrams contributing to self-energy 
as shown in Fig. \ref{PhiPhi_Loop}.

The amplitude for Fig. \ref{PhiPhi_Loop_A} is
\begin{equation}
\mathcal{M}_{nn}^{(1)}=
\left(-2i\frac{e_n^2}{c_n}g^{\mu\nu}\right)
\left(\frac{1}{i}\right)
\int\frac{d^4l}{(2\pi^4)}
\left[\frac{b_0(g_{\mu\nu}-(1-\xi)\frac{l_\mu l_\nu}{l^2})}{l^2}
+\sum_{a=1}^{\infty}\frac{b_a(g_{\mu\nu}+\frac{l_\mu l_\nu}{M_a^2})}{l^2+M_a^2}\right].
\label{selfenergyPPAA}
\end{equation}
The term with the worst UV-divergence is
\begin{equation}
\mathcal{M'}_{nn}\propto\sum_{a=1}^{\infty}\frac{1}{M_a^2}
\int d^4l 
\frac{b_a l^2}{l^2+M_a^2} \propto \Lam^4
\end{equation}
for UV cutoff at $\Lambda$.
In general we also expect UV divergences of lower orders.
It is easy to see that,
in the gauge $\xi = 0$,\footnote{
For our purpose, the gauge $\xi = 0$ is particularly convenient.
In generic gauge, the UV divergence of order $\Lambda^2$ 
for the diagram in Fig. \ref{PhiPhi_Loop_A} is 
actually proportional to
\be
(1-(1-\xi)/4)b_0 + (1-1/4)\sum_{a=1}^{\infty}b_a = 0.
\ee
On the other hand, 
the diagram in Fig. \ref{PhiPhi_Loop_APhi} contributes a factor of 
\be
-\frac{1}{4}(1-(1-\xi))b_0
\ee
to the same divergent term.
The sum of the two diagrams is gauge invariant.
}
UV divergences at various orders are proportional to
\begin{equation}
\sum_{a=1}^{\infty} \frac{1}{M_a^2}b_a \Lambda^4,\quad
\left(b_0+\sum_{a=1}^{\infty} \frac{1}{M_a^2}b_a M_a^2\right)\Lambda^2,
\quad
\sum_{a=1}^{\infty} \frac{1}{M_a^2}b_a M_a^4\log(\Lambda^2).
\label{UVdivorders}
\end{equation}
Therefore, at least three conditions are required
to remove all UV divergences:
\begin{subequations}
\begin{align}
\sum_{a=1}^{\infty}\frac{b_a}{M_a^2} &=0,
\label{PhotoCondition1}\\
\sum_{a=0}^{\infty}b_a &=0,
\label{PhotoCondition2}\\
\sum_{a=1}^{\infty}b_a M_a^2 &=0.
\label{PhotoCondition3}
\end{align}
\label{PhotoCondition}
\end{subequations}

By imposing the condition (\ref{PhotoCondition1}), 
we find something interesting in the vector propagator. 
Calculations below will also be done in the gauge $\xi=0$, 
which is more convenient.
In this gauge, the propagator (\ref{Photopropa}) becomes
\begin{eqnarray}
\Delta_{\mu\nu}(k)
&=&
\frac{b_0(g_{\mu\nu}-\frac{k_\mu k_\nu}{k^2})}{k^2}
+\sum_{a=1}^{\infty}\frac{b_a(g_{\mu\nu}+\frac{k_\mu k_\nu}{M_n^2})}{k^2+M_n^2} \nonumber\\
&=&
\sum_{a=0}^{\infty}\frac{b_a(g_{\mu\nu}-\frac{k_\mu k_\nu}{k^2})}{k^2+M_n^2}
+\sum_{a=1}^{\infty}\frac{b_ak_\mu k_\nu}{k^2+M_n^2}
\left(\frac{1}{M_n^2}+\frac{1}{k^2}\right)
\nonumber\\
&=&
\sum_{a=0}^{\infty}\frac{b_aP_{\mu\nu}}{k^2+M_n^2}
+\frac{k_\mu k_\nu}{k^2}
\left(\sum_{a=1}\frac{b_a}{M_n^2}\right)
\nonumber\\
&=&
\sum_{a=0}^{\infty}\frac{b_aP_{\mu\nu}}{k^2+M_n^2}.
\label{Photopropa2}
\end{eqnarray}
Here 
\begin{equation}
P_{\mu\nu}\equiv g_{\mu\nu}-\frac{k_\mu k_\nu}{k^2}
\end{equation}
is a projection that projects out vectors parallel to $k$.
Therefore, by imposing the condition (\ref{PhotoCondition1}), 
the polarization of a vector propagator is 
perpendicular to its momentum even off-shell!

Now we can rewrite (\ref{selfenergyPPAA}) in the form
\begin{equation}
\mathcal{M}_{nn}^{(1)}=
\left(-2i\frac{e_i^2}{c_n}g^{\mu\nu}\right)\left(\frac{1}{i}\right)
\sum_{a=0}
\int\frac{d^dl}{(2\pi)^d}
\frac{b_a P_{\mu\nu}}{k^2+M_a^2}.
\end{equation}
Integrating it in $d=4-\epsilon$ dimensions gives
\begin{align}
\mathcal{M}_{nn}^{(1)}&=
-2\frac{e_n^2}{c_n}(d-1)
\sum_{a=0}
\int\frac{d^dl}{(2\pi)^d}
\frac{b_a}{k^2+M_a^2} \nonumber\\
&=
-2\frac{e_n^2}{c_n}(3-\epsilon)
\sum_{a=0}
\frac{b_a M_a^2}{(4\pi)^{2-\epsilon /2}}
\left[\frac{-2}{\epsilon}+\gamma-1+\mathcal{O}(\epsilon)\right]
\left[1+\frac{\epsilon}{2}\log(\frac{1}{M_a^2})+\mathcal{O}(\epsilon^2)\right].
\nonumber\\
\end{align}
The coefficient of the $1/\epsilon$ term vanishes 
if the condition (\ref{PhotoCondition3}) is satisfied.
The finite part of the scattering amplitude 
can be obtained by taking the limit $\epsilon\rightarrow 0$,
\begin{equation}
\mathcal{M}_{nn}^{(1)}=
- \frac{3e_n^2}{8\pi^2c_n^2}
\sum_{a=0}
b_a M_a^2
\log(M_a^2).
\label{selfenergyPPAAr}
\end{equation}
In the sum over the index $a$, 
the term for $a = 0$ is ill-defined. 
This is the same infrared divergence in ordinary QED. 
It is resulted from the ignorance of diagrams 
with soft photons ($k=0$). 
A simple way to deal with it is to introduce 
a mass term $m_\gamma$ to the photon 
as an infrared regulator 
and take the limit $m_\gamma\rightarrow 0$ at the end. 
The final result is that the $a=0$ term has no contribution.

Now we calculate the amplitude for Fig. \ref{PhiPhi_Loop_APhi}. 
It is
\begin{align}
\mathcal{M}_{nn}^{(2)} &=
\left(\frac{ie_n}{c_n}\right)^2\left(\frac{1}{i}\right)^2
\int\frac{d^4l}{(2\pi)^4}
(l+2p)^\mu
\frac{c_n}{(l+p)^2+m_n^2}
(l+2p)^\nu
\sum_{a=0}\left(\frac{b_a P_{\mu\nu}(l)}{l^2+M_a^2}\right)
\nonumber
\\
&=
e_n^2(2p)^\mu(2p)^\nu
\frac{1}{c_n}\sum_{a=0}b_a
\int\frac{d^dl}{(2\pi)^d}
\frac{1}{(l+p)^2+m_n^2}
\frac{P_{\mu\nu}(l)}{l^2+M_a^2}.
\label{selfenergyPPAs}
\end{align}
Multiplying the numerator and denominator in (\ref{selfenergyPPAs}) 
by $l^2$ and using Feynman's parameters,
we can compute the essential part of the integral above as
\begin{equation}
\int\frac{d^dl}{(2\pi)^d}
\frac{l^2 P_{\mu\nu}(l)p^\mu p^\nu}{l^2(l^2+M_a^2)((l+p)^2+m_n^2)}
=
\int dF_3
\int\frac{d^dl}{(2\pi)^d}
\frac{N}{(q^2+\Delta)^3},
\end{equation}
where 
\begin{align}
\int dF_3&\equiv 2\int_0^1\int_0^1\int_0^1 
d\alpha_1 d\alpha_2 d\alpha_3
\delta(\alpha_1+\alpha_2+\alpha_3-1), \\
q&=\alpha_1l+\alpha_2l+\alpha_3(l+p)=l+\alpha_3 p, \\
\Delta
&=
\alpha_3(1-\alpha_3)p^2+\alpha_2M_a^2+\alpha_3m_n^2, \\
N
&= q^2p^2-d^{-1}q^2p^2.
\end{align}
Again we apply dimensional regularization
to define the integral
\begin{equation}
\sum_{a=0}b_a\int\frac{d^dl}{(2\pi)^d}
\frac{q^2}{(q^2+\Delta)^3}
\approx
\sum_{a=0}b_a
\left[\frac{2}{\epsilon}-\log\Delta-\gamma
+\mathcal{O}(\epsilon)\right]
\frac{1}{(4\pi)^{2}}.
\end{equation}
The $1/\epsilon$ term vanishes with the condition
(\ref{PhotoCondition2}).
The finite part is
\begin{equation}
\mathcal{M}_{nn}^{(2)}=
-\frac{6e_n^2}{c_n}
\int_0^1 \prod_{i=1}^3 d\alpha_i
\delta(\alpha_1+\alpha_2+\alpha_3-1)\sum_{a=0}b_a
\log[\alpha_3(1-\alpha_3)p^2+\alpha_2M_a^2+\alpha_3m_n^2].
\label{selfenergyPPAr}
\end{equation}

\subsubsection{Choice of parameters}

We can construct a solution of parameters to 
all the conditions (\ref{PhotoCondition}) 
in a way similar to (\ref{example1}). 
Because $M_0^2=0$, 
for convenience we rewrite (\ref{PhotoCondition}) in the form
\begin{subequations}
\begin{align}
&\sum_{a=1}^{\infty}\frac{b_a}{M_a^2}=0,
\\
&\sum_{a=1}^{\infty}b_a=-b_0,
\label{b0value}
\\
&\sum_{a=1}^{\infty}b_a M_a^2=0.
\end{align}
\label{PhotoCondition_2}
\end{subequations}
A special class of solutions of the parameters is
\begin{subequations}
\begin{align}
b_a&=\Bigl[
a+1+x_1(a+2)(a+1)+x_2(a+1)(a+2)(a+3)\Bigr]e^{za}
\qquad (a\ge 1, x_i\ge 0),
 \\
M^2_a&=e^{\a a} \qquad (a\ge 1, \a > 0).
\end{align}
\end{subequations}
The value of $b_0$ will be determined by (\ref{b0value}), 
and $M_0$ is $0$.
In addition to the reality condition $M^2_a > 0$,
we also need $b_a > 0$ for all $a = 0, 1, 2, \cdots$
for the sake of unitarity.
The values of $z$ and $\a$ will be set to be positive numbers.

That the solution above satisfies all conditions
can be seen as follows.
Let $\rho\equiv e^{z+\a r},\;r=\pm 1$. 
We compute $\sum b_a M^{2r}_a$ assuming $\rho < 1$
and then analytically continue $\rho$ back to $\rho>1$.
We get
\begin{subequations}
\begin{align}
\sum_{a=1}^\infty b_a M_a^{2r}
&=\frac{d}{d\rho}\left(\frac{1}{1-\rho}\right)
+x_1\frac{d^2}{d\rho^2}\left(\frac{1}{1-\rho}\right)
+x_2\frac{d^3}{d\rho^3}\left(\frac{1}{1-\rho}\right) \\
&=\frac{1}{\xi^2}+\frac{x_1}{\xi^3}+\frac{x_2}{\xi^4}
\equiv h(\xi),
\label{choice3}
\end{align}
\end{subequations} 
where $\xi\equiv\frac{1}{1-\rho}$, 
which is negative definite when $\rho>1$.
By this method, 
we have sufficient parameters $\{x_1,x_2\}$ 
to fix the root of $\xi$ at desired positions. 
Defining the roots as 
\begin{equation}
-|\xi_1|=\frac{1}{1-e^{z-\a}}, \qquad
-|\xi_2|=\frac{1}{1-e^{z}}, \qquad
-|\xi_3|=\frac{1}{1-e^{z+\a}}.
\end{equation}
We see that $|\xi_3|>|\xi_2|>|\xi_1|>0$.
The corresponding values of $x_1$ and $x_2$ 
can be found by simply comparing the coefficients 
with the following equation
\begin{equation}
\xi^4 h(\xi)=c(\xi+|\xi_1|)(\xi+|\xi_3|),
\label{findx_2}
\end{equation}
where $c$ is an arbitrary positive real parameter.
Both $x_1$ and $x_2$ are guaranteed to be positive 
because no negative coefficient appears in (\ref{findx_2}). 
Furthermore, $h(-|\xi_2|)$ is negative 
so that we let
\begin{equation}
b_0=-h(-|\xi_2|)>0.
\end{equation}

The finite part of a diagram is free from UV divergence, 
but it is not necessarily finite because it involves an infinite series.
Ignoring the integration of Feynman parameters, 
$\mathcal{M}^{(1)}_{nn}$ and $\mathcal{M}^{(2)}_{nn}$ 
are both of the form
\begin{equation}
\sum_{a_1\cdots a_I}b_{a_1}\cdots b_{a_I}
\Delta^r\log\Delta.
\label{finitesum}
\end{equation}
The magnitude of $\Delta$ is roughly of the same order as $M^2_a$, 
which grows exponentially with $a$,
so $\log\Delta$ grows like $a$ when $a$ increases.
On the other hand, 
$b_a$ decays exponentially with $a$ for $z<0$. 
For $z$ sufficiently negative, the series converges. 
The next step is to analytically continue $z$ to $z_0>0$. 
This might lead to new divergence.
But it is free from divergences as long as 
\begin{equation}
m z_0-n\a\neq 0\qquad\forall m,n\in \mathbb{Z}.
\label{finiteness}
\end{equation}
That is, we want the ratio $z_0/\a$ to be an irrational number.
This condition can be understood as the following: 
this choice of parameters (\ref{PhotoCondition_2}) is based on 
the analytic continuation of geometric series. 
But the series can not be defined even by analytic continuation
when the ratio between successive terms is equal to one.
To prevent this for a generic series of the form (\ref{finitesum}), 
the condition
\begin{equation}
e^{mz_0-na}\neq 1\qquad\forall m,n\in \mathbb{Z}
\end{equation}
is sufficient.

\subsubsection{Self-energy of $A_{a,\mu}$}

\begin{figure}[!ht]
\centering
\subfigure[] {
\label{AA_Loop_PhiPhi_2}
\includegraphics[scale=0.5]{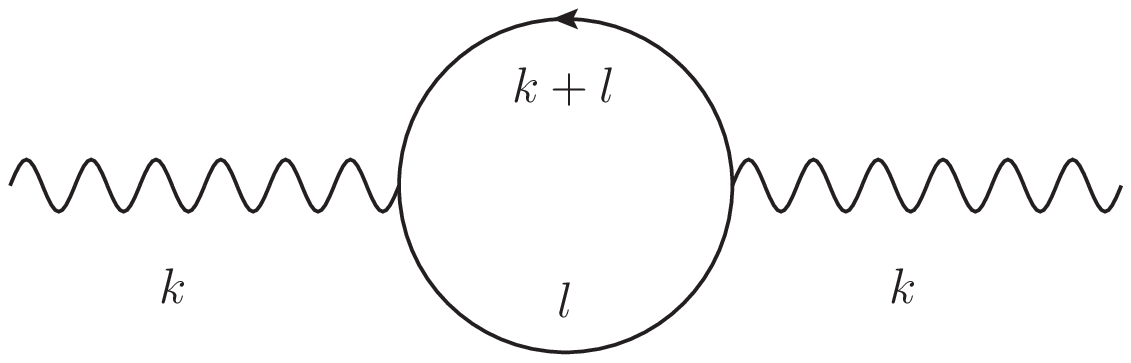}
}
\hspace{1cm}
\subfigure[] {
\label{AA_Loop_Phi}
\includegraphics[scale=0.5]{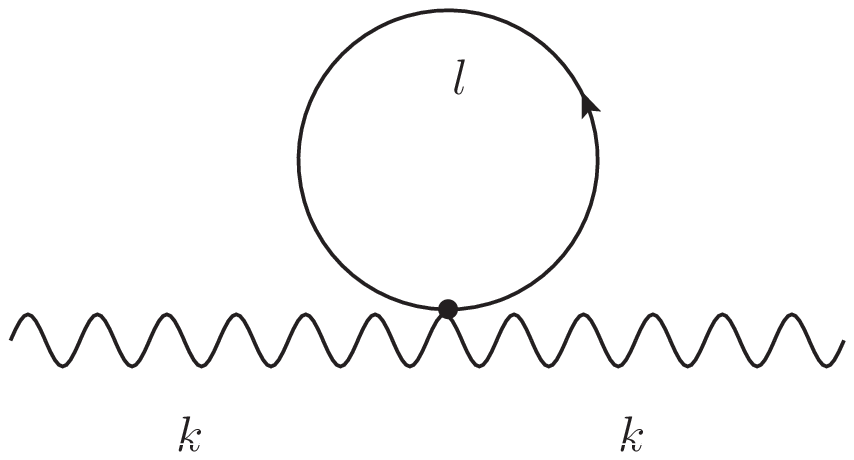}
}
\caption{}
\label{AA_Loop_PhiPhi_2+15}
\end{figure}

Now we calculate the vector field's self-energy. 
The one-loop corrections to the vector propagator 
are shown in Fig \ref{AA_Loop_PhiPhi_2+15}.

The sum of these two diagrams is
\begin{align}
\Pi^{\mu\nu}(k)&=
\left(\frac{1}{i}\right)^2
\sum_n \left(\frac{ie_n}{c_n}\right)^2\int\frac{d^4l}{(2\pi)^4}
(2l+k)^\mu\frac{c_n}{l^2+m_n^2}(2l+k)^\nu \frac{c_n}{(l+k)^2+m_n^2}
\nonumber\\
&+
(-2i)g^{\mu\nu}\left(\frac{1}{i}\right)
\sum_n\frac{e_n^2}{c_n}\int\frac{d^4l}{(2\pi)^4}
\frac{c_n}{l^2+m_n^2}\nonumber\\
&=
\sum_ne_n^2\int\frac{d^4l}{(2\pi)^4}
(2l+k)^\mu\frac{1}{l^2+m_n^2}(2l+k)^\nu \frac{1}{(l+k)^2+m_n^2}
\nonumber\\
&+
(-2)g^{\mu\nu}
\sum_ne_n^2\int\frac{d^4l}{(2\pi)^4}
\frac{1}{l^2+m_n^2}.
\label{selfenergyAA}
\end{align}
As all factors of $c_n$'s cancel,
the UV divergence can only be cancelled
by tuning the coupling constants $e_n$.

The first term in (\ref{selfenergyAA}) can be written in the form
\begin{align}
\Pi^{(1)\mu\nu}(k)&=
\sum_ne_n^2\int\frac{d^4l}{(2\pi)^4}
\frac{(2l+k)^\mu(2l+k)^\nu}{[(l+k)^2+m_n^2][l^2+m_n^2]}
\nonumber\\
&=
\int_0^1 d\alpha
\sum_ne_n^2\int\frac{d^4q}{(2\pi)^4}
\frac{N^{\mu\nu}}{(q^2+\Delta_{nn})^2},
\label{selfenergyAAP}
\end{align}
where
\begin{subequations}
\begin{align}
q&=\alpha (l+k)+(1-\alpha)l=l+\alpha k,\\
\Delta_{nn}&=m_n^2+\alpha(1-\alpha)k^2,\\
N^{\mu\nu}
&= g^{\mu\nu}q^2+(1-2\alpha)^2 k^\mu k^\nu.
\end{align}
\end{subequations}
In $d=4-\epsilon$ dimensions,
(\ref{selfenergyAAP}) can be calculated
\begin{align}
\Pi^{(1)\mu\nu}(k)
&=
\int_0^1d\alpha\sum_n2e_n^2
\frac{g^{\mu\nu}\Delta_{nn}}{16\pi^2}
\left[\frac{-2}{\epsilon}+\gamma-1+\mathcal{O}(\epsilon)\right]
\left[1+\frac{\epsilon}{2}\log(\frac{1}{\Delta_{nn}})
+\mathcal{O}(\epsilon^2)\right] \nonumber\\
&\quad+
\int_0^1 d\alpha\sum_ne_n^2
\frac{(1-2\alpha)^2 k^\mu k^\nu}{16\pi^2}
\left[\frac{2}{\epsilon}-\log\Delta_{nn}-\gamma
+\mathcal{O}(\epsilon)\right].
\label{selfenergyAAP2}
\end{align}
For the $1/\epsilon$ terms in (\ref{selfenergyAAP2}) to vanish, 
we need two conditions
\begin{subequations}
\begin{align}
&\sum_{n=0}^{\infty} e_n^2=0, \\
&\sum_{n=0}^{\infty} e_n^2 m_n^2=0.
\label{conditione2m2}
\end{align}
\label{conditionem}
\end{subequations}
These are reminiscent of those in (\ref{condition}), 
with $c_n$'s replaced by $e_n$'s.

The finite part of this amplitude is 
\begin{align}
\Pi^{(1)\mu\nu}(k)=
\frac{g^{\mu\nu}}{8\pi^2}
\int_0^1d\alpha\sum_{n=0}^{\infty} e_n^2
\Delta_{nn}\log\Delta_{nn}
-
\int_0^1 d\alpha\sum_{n=0}^{\infty} e_n^2
\frac{(1-2\alpha)^2 k^\mu k^\nu}{16\pi^2}
\log\Delta_{nn}.
\label{selfenergyAAPr}
\end{align}

The second term in (\ref{selfenergyAA}) is
\begin{equation}
\Pi^{(2)\mu\nu}(k)=
(-2)g^{\mu\nu}
\sum_ne_n^2\int\frac{d^4l}{(2\pi)^4}
\frac{1}{l^2+m_n^2}.
\label{selfenergyAAPP}
\end{equation}
The calculation is straightforward.
UV-divergences cancel due to (\ref{conditionem}),
and the finite part is
\begin{equation}
\Pi^{(2)\mu\nu}(k)=
-\frac{g^{\mu\nu}}{8\pi^2}
\sum_{n=0}^{\infty}
e_n^2m_n^2
\log({m_n^2}).
\label{selfenergyAAPPr}
\end{equation}

Integrating over the Feynman parameter in 
(\ref{selfenergyAAPr}) and adding it to (\ref{selfenergyAAPPr}),
we compute the self-energy of $A_{a, \mu}$ as
\begin{equation}
\Pi^{\mu\nu}(k)=\frac{g^{\mu\nu}k^2-k^\mu k^\nu}{8\pi^2}
\sum_{n=0}^{\infty} e_n^2
\left[
\frac{\sqrt{\frac{(k^2+4m_n^2)^3}{3k^2}}
\tanh^{-1}\left(\sqrt{\frac{k^2}{k^2+4m_n^2}}\right)}{k^2}
+\frac{\log m_n}{3}
\right].
\label{selfenergyAAr}
\end{equation}

In order for the $U(1)$ gauge symmetry to be valid at the quantum level, 
the photon must remain massless against quantum corrections. 
The mass correction of the photon can be derived from its self-energy. 
Eq. (\ref{selfenergyAAr}) only provides its finiteness. 
We need to check that the mass correction vanishes. 

At one loop,
the vector propagator is given by
\begin{equation}
\mathbf{\Delta}^\mu_\nu(k)=
\Delta^\mu_\nu
+\Delta^\mu_{\alpha_1}\Pi^{\alpha_1}_{\alpha_2}\Delta^{\alpha_2}_\nu
+\Delta^\mu_{\alpha_1}
  \Pi^{\alpha_1}_{\alpha_2}\Delta^{\alpha_2}_{\alpha_3}
  \Pi^{\alpha_3}_{\alpha_4}\Delta^{\alpha_4}_\nu+\cdots.
\label{photonmasscorrection}
\end{equation}
Let us define a scalar function $\Delta_0(k)$ 
in the photon propagator for convenience
\begin{equation}
\Delta^{\mu\nu}_0(k)=P^{\mu\nu}\Delta_0(k),
\end{equation}
where
\begin{equation}
\Delta_0(k)\equiv\frac{b_0}{k^2}.
\end{equation}
Similarly, we rewrite (\ref{selfenergyAAr}) as
\begin{equation}
\Pi^{\mu\nu}(k)=\frac{P^{\mu\nu}}{8\pi^2}k^2\Pi(k),
\label{selfenergyAAr2}
\end{equation}
where
\begin{equation}
\Pi(k)\equiv
\sum_{n=0}^{\infty}\frac{e_n^2}{8\pi^2}
\left[
\frac{\sqrt{\frac{(k^2+4m_n^2)^3}{k^2}}
\tanh^{-1}\left(\sqrt{\frac{k^2}{k^2+4m_n^2}}\right)}{3k^2}
+\frac{\log m_n}{3}
\right].
\label{selfenergyAArscalar}
\end{equation}
Note that only the massless propagator $\Delta^{\mu\nu}_0(k)$ 
appears in internal lines. 
The diagram in Fig. \ref{AA_Propagator_Modify} 
with $n \neq 0$ is considered a two-loop correction.
\begin{figure}[t]
\centering
\scalebox{0.5}{
  \begin{picture}(812,130) (39,-175)
    \SetWidth{1.0}
    \SetColor{Black}
    \Photon(96,-110)(254,-110){7.5}{7}
    \Photon(340,-110)(498,-110){7.5}{7}
    \Photon(600,-110)(784,-110){7.5}{7}
    \Text(158,-158)[lb]{\LARGE{\Black{$A_0$}}}
    \Text(400,-158)[lb]{\LARGE{\Black{$A_a$}}}
    \Text(688,-158)[lb]{\LARGE{\Black{$A_0$}}}
    \GOval(290,-110)(50,50)(0){0.882}
    \GOval(548,-110)(50,50)(0){0.882}
    \Text(36,-122)[lb]{\LARGE{\Black{$\sum \quad( $}}}
    \Text(804,-126)[lb]{\LARGE{\Black{$)^{n}$}}}
  \end{picture}
}
\caption{}
\label{AA_Propagator_Modify}
\end{figure} 
The infinite series in (\ref{photonmasscorrection}) can be 
largely simplified as
\begin{align}
\mathbf{\Delta}^\mu_{0,\nu}(k)&=
P^\mu_\nu
(\Delta_0(k)+\Delta_0k^2\Pi(k)\Delta_0(k)
+\Delta_0k^2\Pi(k)\Delta_0(k)k^2\Pi(k)\Delta_0(k) + \cdots) 
\nonumber\\
&=
P^\mu_\nu
\frac{b_0}{k^2-b_0 k^2\Pi(k)} .
\end{align}
In the neighborhood of the pole $k^2=0$, the propagator is approximated by 
\begin{equation}
\Delta^\mu_{0,\nu}(k)\simeq
P^\mu_\nu(0)
\frac{b_0}{k^2-b_0[\lim_{k^2\rightarrow 0}k^2\Pi(k)]},
\end{equation}
and so the mass correction is
\begin{equation}
\delta m_\gamma^2=\lim_{k^2\rightarrow 0}
b_0k^2\Pi(k),
\end{equation}
Using (\ref{selfenergyAArscalar}) and the condition (\ref{conditione2m2}),
we compute the mass correction as
\begin{align}
\delta m_\gamma^2&=
\lim_{k^2\rightarrow 0}
\sum_{n=0}\frac{e_n^2}{24\pi^2}
\left[
\sqrt{\frac{(k^2+4m_n^2)^3}{k^2}}
\tanh^{-1}\left(\sqrt{\frac{k^2}{k^2+4m_n^2}}\right)
+k^2\log m_n
\right] \nonumber\\
&=
\sum_{n=0}\frac{e_n^2}{24\pi^2}\times
4m_n^2 \nonumber\\
&=0.
\end{align}
We find that the mass correction of photon is indeed zero.

\subsection{Generic Feynman diagrams}

Here we discuss the finiteness of generic Feynman diagrams. 
We will show that the power of superficial divergence $D$ 
of all Feynman diagrams are negative.

Define $L$ as the number of loops, 
$I_A$/$E_A$ as the number of vector internal/external lines, 
$I_\phi$/$E_\phi$ as the number of scalar internal/external lines, 
$V_3$ as the number of the vertices $\phi\phi A$, 
$V_4$ as the number of the vertices $\phi\phi AA$.
These parameters satisfy several algebraic relations 
for all diagrams.
First, we have
\begin{equation}
L=I_\phi+I_A-V_3-V_4+1.
\label{Loopandvertexandinternal}
\end{equation} 
This equality states that the number of free momenta 
(on the loops)
is equal to the number of internal lines 
minus the constraints due to energy-momentum conservation.
For each vertex, 
there is a constraint on energy-momentum conservation,
but that for the external momenta is irrelevant for
the loop momenta.

Each vertex $\phi\phi A$ has only one leg of vector, and each vertex $\phi\phi AA$ has two. Therefore,
\begin{equation}
2V_4+V_3=2I_A+E_A.
\label{VVI_AE_A}
\end{equation} 
Similarly, both vertices $\phi\phi A$ and $\phi\phi AA$ have two legs of scalar,
\begin{equation}
2V_4+2V_3=2I_\phi+E_\phi.
\label{VVI_PE_P}
\end{equation}

In the previous subsections, 
we saw that vector internal lines can always 
be treated as a superposition of all vector particles, 
while the scalar internal lines cannot. 
The effective vector propagator is (\ref{Photopropa2})
\begin{align}
\Delta_{\mu\nu}&=\sum_{a=0}^{\infty}
\frac{b_a P_{\mu\nu}}{k^2+M_a^2} \nonumber\\
&=P_{\mu\nu}
\left(
\frac{\sum_a b_a}{k^2}-\frac{\sum_a b_a M_a^2}{k^4}
+\frac{\sum_a b_a M_a^4}{k^6}+\cdots
\right)\nonumber\\
&=\sum_{a=0}^{\infty}\frac{b_a P_{\mu\nu }M_a^4}{k^4(k^2+M_a^2)},
\end{align}
which goes like $1/k^6$ at high energies.
The power of superficial UV divergence of an arbitrary diagram is 
\begin{equation}
D=4L+V_3-2I_\phi-6I_A.
\label{SuperficialD}
\end{equation}
Each loop integral contributes a 4 dimensional integral 
of energy-momentum, 
the propagator of $A$ a factor of $1/k^6$
and the propagator of $\phi$ a factor of $1/k^2$ at high energies. 
Furthermore each vertex $\phi\phi A$ 
contributes a linear power of momentum due to the coupling.

Combining (\ref{Loopandvertexandinternal}), 
(\ref{VVI_AE_A}), (\ref{VVI_PE_P}) and (\ref{SuperficialD}), 
we get
\begin{equation}
D=4-4I_A-E_A-E_\phi.
\end{equation}
The superficial divergence is bounded from above, 
as in $\phi^4$ theory. 
With the assumption that the superficial divergence 
reflects the real power of divergence,
UV finiteness is guaranteed if
\begin{equation}
I_A > 1.
\end{equation}
When $I_A = 1$, 
$D$ is non-negative only if
there is no external lines of $A$ nor $\phi$.
The only such possibility is then a vacuum diagram 
composed of a single loop of $A$.
This diagram actually has $D = -2$
and is free from UV divergence.
When $I_A = 0$, that is, 
all vectors are external, 
all loops are composed only of scalar internal lines.
A loop of scalar is divergent only when 
it is composed of no more than two scalar internal lines.
These cases are already studied earlier
when we computed the self-energy of the vector field
(Fig. \ref{AA_Loop_PhiPhi_2+15}).
They are also UV finite.
Thus we believe that all diagrams of the theory 
are UV finite.

\section{Discussion}
\label{Discussion}

\subsection{Self interactions of $\phi$}

In the above we assumed that the self interactions of 
the scalars vanish. 
That is, 
\begin{equation}
V(\sum_n \phi_n^{\dag}\phi_n) = 0.
\end{equation}
We briefly discuss here the situation when $V \neq 0$.

Because of the global symmetry (\ref{phiphase}),
a scalar propagator is not always superposed over all $\phi_n$, 
and so it has no suppression 
due to the conditions (\ref{condition}) at high energies.
Even in 4 dimensions, 
the conditions (\ref{condition}) no longer ensure
the finiteness of a loop consisting of only scalar interaction vertices.

Actually we have already seen the effect of 
the global symmetry (\ref{phiphase}) above.
For example, in the diagram for self-energy of $\phi_n$
(Fig. \ref{PhiPhi_Loop}),
the index of the scalar internal line is fixed by the external lines. 
Similarly, in the self-energy diagram Fig. \ref{AA_Loop_PhiPhi_2},
the index $n$ of the two scalar propagators are not independent,
and the UV divergence can not be cancelled 
by tuning the values of $c_n$.
\begin{figure}[t]
\centering
\scalebox{0.5}{
  \begin{picture}(197,199) (46,5)
    \SetWidth{2.0}
    \SetColor{RoyalPurple}
    \Line[arrow,arrowpos=0.5,arrowlength=6.667,arrowwidth=2.667,arrowinset=0.2](48,7)(96,55)
    \Line[arrow,arrowpos=0.5,arrowlength=6.667,arrowwidth=2.667,arrowinset=0.2](96,135)(48,183)
    \SetColor{PineGreen}
    \Line[arrow,arrowpos=0.5,arrowlength=6.667,arrowwidth=2.667,arrowinset=0.2](176,55)(224,7)
    \Line[arrow,arrowpos=0.5,arrowlength=6.667,arrowwidth=2.667,arrowinset=0.2](224,183)(176,135)
    \Arc[arrow,arrowpos=0.5,arrowlength=6.667,arrowwidth=2.667,arrowinset=0.2](136,95)(56.569,45,315)
    \SetColor{RoyalPurple}
    \Line[arrow,arrowpos=0.5,arrowlength=6.667,arrowwidth=2.667,arrowinset=0.2](96,55)(130,90)
    \Line[arrow,arrowpos=0.5,arrowlength=6.667,arrowwidth=2.667,arrowinset=0.2](140,100)(176,135)
    \Line[arrow,arrowpos=0.5,arrowlength=6.667,arrowwidth=2.667,arrowinset=0.2](176,55)(96,135)
    \Arc[arrow,arrowpos=0.5,arrowlength=6.667,arrowwidth=2.667,arrowinset=0.2,clock](136,95)(56.569,45,-45)
    \Text(64,7)[lb]{\LARGE{\Black{$\phi_n$}}}
    \Text(200,90)[lb]{\LARGE{\Black{$\phi_n$}}}
    \Text(64,180)[lb]{\LARGE{\Black{$\phi_n$}}}
    \Text(128,110)[lb]{\LARGE{\Black{$\phi_n$}}}
    \Text(192,7)[lb]{\LARGE{\Black{$\phi_m$}}}
    \Text(185,180)[lb]{\LARGE{\Black{$\phi_m$}}}
    \Text(88,87)[lb]{\LARGE{\Black{$\phi_m$}}}
    \Text(130,42)[lb]{\LARGE{\Black{$\phi_m$}}}
  \end{picture}
}
\caption{Every internal line is fixed to be labelled by 
$m$ (green line) or $n$ (purple line), 
and so there is no superposition of propagators
for better UV behavior.
}
\label{tetra-diagram}
\end{figure}
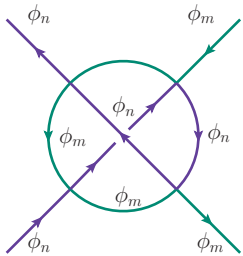
As we have seen earlier, 
these diagrams are UV-finite if 
the conditions (\ref{PhotoCondition}) 
and (\ref{conditionem}) are satisfied.
But what happens if the propagator of $A$ is replaced 
by a propagator of $\phi_n$, 
when the vertices are replaced by scalar self interactions?

It turns out that there are Feynman diagrams 
(e.g. Fig. \ref{tetra-diagram}) with logarithmic divergence 
which can not be removed by any condition on the coefficients
because all the indices are fixed by the external lines.
Therefore the assumption $V = 0$
is a necessity for the sake of UV convergence.

\subsection{Spinor electrodynamics}

It is not difficult to generalize this framework to fermions. 
The Lagrangian is of the form
\begin{eqnarray}
\mathcal{L}&=&\sum_n\frac{-1}{c_n}\biggl(
i\Psib_n \dels \Psi_n 
+e\Psib_n \gamma^\mu A_\mu \Psi_n - m_n\Psib_n \Psi_n
\biggr) \nonumber\\
&&-\sum_a\frac{1}{b_a}\left(\frac{1}{4}F_a^{\m\n}F_{a,\m\n}
+\frac{1}{2}M_a^2 A_a^\m A_{a,\m}\right).
\label{LagrangianF}
\end{eqnarray}
The Feynman rule is even simpler as 
there is a single three-point interaction for the fermion and vector.

All discussions above can be repeated easily. 
All results are also similar to the scalar theory.
The constraints on fermion masses are slightly modified
\begin{align}
&\sum_n c_n m_n^r=0,\qquad r=0,1,2,3\quad
\text{(from fermion self interactions)},
\label{fermicondch3} \\
&\sum_n e^2_n m_n^r=0,\qquad r=0,1,2,3\quad
\text{(from fermion-vector interactions)}.
\label{fermicondch3_2}
\end{align}
Eq.(\ref{fermicondch3}) is the same as (\ref{fermicond}),
and (\ref{fermicondch3_2}) is the fermionic version 
of (\ref{conditionem}).

\subsection{Renormalization}

Massive vector field theories, 
as well as $\phi^n$ theories for $n>4$ in four dimensions, 
have been considered non-renormalizble. 
These theories require infinite number of counter terms and 
they contain an infinite number of free parameters.

In our theory, 
we also have an infinite number of parameters, 
but they are not totally free. 
To satisfy finiteness conditions such as (\ref{condition}), 
the possible choices of these parameters are restricted. 
Furthermore, by calculating the explicit form of 
an amplitude by analytic continuation, 
we can also compare the calculation with experimental data 
and determine which choice of parameters is correct.

Our recipe can also be viewed as merely 
a new method of regularization. 
After removing all UV-divergence, 
we can talk about how coupling constants run with energy, 
and we can still apply the renormalization procedure to our theory.
It will be very interesting to study the properties 
of the renormalization group flow of our models.

\subsection{Future works}

An important generalization of our results is to extend it to 
non-abelian gauge theories. 
For Abelian gauge symmetry, 
a renormalizable, unitary perturbative theory 
of the massive vector boson
can be constructed with the help of a Stueckelberg field.
But non-Abelian gauge symmetry is much harder to deal with.
Despite years of efforts,
Higgs mechanism remains the only way to 
maintain unitarity and renormalizability for 
the existence of massive vector bosons.
It will be very interesting to see whether our idea 
can be generalized to deal with non-Abelian theories,
and to construct an alternative mechanism to give masses 
to vector bosons.

\section*{Acknowledgment}

The authors thank Chuan-Tsung Chan, Jiunn-Wei Chen, 
Hsien-chung Kao and Yeong-Chuan Kao for helpful discussions. 
This work is supported in part by the National Science Council,
and the National Center for Theoretical Sciences, Taiwan, R.O.C.

\end{document}